\documentclass[12pt,preprint]{aastex}

\newcommand{\msol}{\,$M_{\odot}$}

\begin{document}

\title{A multiwavelength radial velocity search for planets around the 
brown dwarf LP 944-20}

\author{E.L. Mart\'{\i}n\altaffilmark{1,2}, 
E. Guenther\altaffilmark{3}, M.R. Zapatero Osorio\altaffilmark{4}, 
H. Bouy\altaffilmark{1},  and R. Wainscoat\altaffilmark{5}} 

\altaffiltext{1}{Instituto de Astrof\'\i sica de Canarias, La Laguna, 
Tenerife 38200, Spain} 
\altaffiltext{2}{University of Central Florida, Department of Physics, PO Box 162385, 
Orlando, FL 32816, USA} 
\altaffiltext{3}{Thuringer Landessternwarte Tautenburg, 07778 Tautenburg, 
Germany} 
\altaffiltext{4}{LAEFF-INTA, P.O. Box 50727, E-28080, Madrid, Spain} 
\altaffiltext{5}{Institute of Astronomy. University of Hawaii, 
2680 Woodlawn Drive. Honolulu, HI 96822, USA}


\begin{abstract}

The nearby brown dwarf LP\,944$-$20 has been monitored for radial 
velocity variability at optical and near-infrared wavelengths using 
the VLT/UVES and the Keck/NIRSPEC spectrographs, respectively. 
The UVES radial velocity data obtained over 14 nights spanning 
a baseline of 841 days shows significant variability with an amplitude 
of 3.5 km\,s$^{-1}$. The periodogram analysis of the UVES data indicates 
a possible period  between 2.5 hours and 3.7 hours, which is likely due  
to the rotation of the brown dwarf. However, the NIRSPEC data obtained 
over 6 nights shows an rms dispersion of only 0.36 km\,s$^{-1}$ and 
do not follow the periodic trend. These 
results indicate that the variability seen with UVES is 
likely to be due to rotationally modulated inhomogeneous 
surface features. We suggest that future planet 
searches around very low-mass stars and 
brown dwarfs using radial velocities will be better 
conducted in the near-infrared than in the optical.  
 
\end{abstract}


\section{Introduction}

Brown dwarfs (hereafter BDs) are objects with very low-masses  
($M\,<\,0.075$\msol) that develop degenerate cores before being 
able to settle on the stellar main-sequence \citep{kumar63}. 
Large numbers of BDs have been identified in the field, 
in open clusters and in 
star-forming regions. These substellar-mass objects could be 
as numerous as the known stars \citep{bejar01,chabrier03}. 

The detection of strong emission lines in young BDs, as well as 
near-infrared, mid-infrared and millimiter emission exceeding 
those expected for normal photospheres, strongly indicate that 
young BDs have disks where planets may form  \citep{barrado03,klein03, 
luhman05}. Moreover, mid-infrared spectroscopic 
observations obtained with Spitzer/IRS indicate dust evolution 
in the discs around young BDs \citep{apai05}.  

One of the first obvious targets to search for planets 
around BDs is LP\,944$-$20 (BRI~B0337-3535; LEHPM~3451; 
2MASS~J03393521-3525440; $J_{\rm 2MASS}=$10.72, $(J-K)_{\rm 2MASS}=$1.18) 
because it is nearby (d=5.0 pc, \citep{tinney98}) 
and young (age$\sim$320~Ma, \citep{ribas03}). 
This BD was first spotted over 30 years ago \citep{luyten75} 
as a dim high proper-motion 
red star, and its substellar nature was revealed through the  
spectroscopic detection of lithium, a trademark of 
BDs \citep{magazzu93, tinney98}. LP\,944$-$20 has a relatively early spectral class 
for a BD (dM9, \citep{martin99}) because of its young age, 
and has a low-level of chromospheric 
and X-ray coronal activity \citep{marbouy02}, 
although occassional flares have been reported \citep{rutledge}. 
Nevertheless, LP944-20 was detected as a radio source  
suggesting the presence of strong magnetic activity \citep{berger01}.   

In this paper we present a search for giant planets around  
LP\,944$-$20. We have monitored the radial velocity (hereafter RV) of 
this BD over 20 nights at optical and near-infrared wavelengths. 
We find that the RV data obtained in the optical is variable with 
an amplitude of 3.5 km\,s$^{-1}$ and a period between 2.5 and 3.7 hours. 
However the RV 
data measured in our near-infrared spectra have an rms dispersion of 
0.36 km\,s$^{-1}$ and does not follow the periodic pattern seen in the optical. 
We discuss the implications of these results for RV searches for planets 
around brown dwarfs.

\section{Observations and results}

The visible spectroscopic observations of LP\,944$-$20 were 
obtained with the Uv-Visual Echelle Spectrograph (UVES) on the 
VLT Unit telescope 2 (KUEYEN) in service mode. The instrumental 
configuration was the same as described in \citep{guenther03}. 
This setting covers simultaneously the regions from 667.0 nm through 
854.5 nm and from 864.0 nm to 1040.0 nm. It includes the telluric 
bands between 686.0 nm and 693.0 nm and between 760.0 nm and 770.0 nm. 
The telluric lines were used as secondary wavelength reference because 
they are know to be stable within 15 m\,s$^{-1}$ \citep{bal82,Smith82,Caccin85}. 
Six UVES RV measurements of LP\,944$-$20 have been reported in the literature  
and a possible RV variability of this object has been noted \citep{guenther03}. 
In this paper we report nine additional UVES RV points. All of them 
were computed in the same way as described in detail in previous work 
\citep{guenther03}. 

\begin{figure}
\epsscale{.70}
\plotone{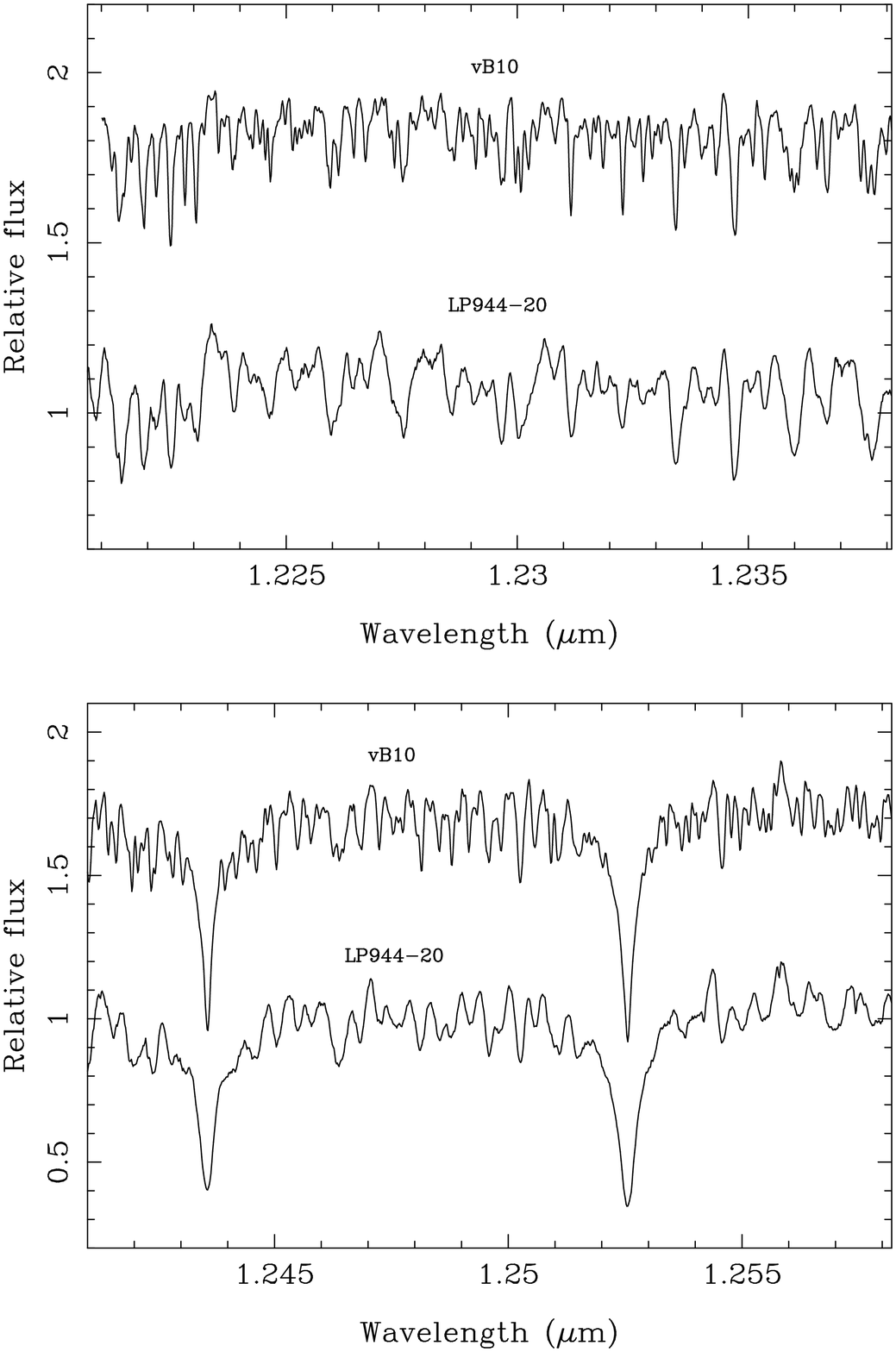}
\caption{\label{combo} 
Average NIRSPEC spectra for our target (LP\,944$-$20) and the RV 
template (VB10). The spectra have been shifted to vacuum 
wavelength and are offset in the vertical axis. 
The upper panel shows a NIRSPEC order dominated by FeH 
molecular lines, whereas the lower panel shows an order which is 
dominated by a KI doublet and  FeH molecular lines. Both of them 
were used to measure the radial velocity of this brown dwarf.} 
\end{figure}

Our infrared spectroscopic observations of LP\,944$-$20 were 
obtained with NIRSPEC on the Keck II telescope. They have already been 
described  in the context of the measurement 
of rotational broadening \citep{zapatero06}. 
We obtained one more NIRSPEC RV datapoint 
in January 2006 which was not included in the rotational velocity paper. 
The data reduction was identical to that of the previous spectra. 
Near-infrared measurements of the heliocentric RV of LP\,944$-$20 
were computed via cross 
correlation of the NIRSPEC data with spectra of the field vB\,10 
(GJ\,752B, dM8), which was observed with very similar instrumental 
configuration in 2001 June 15 and 2001 November 02 (see 
Fig.~\ref{combo}). 
From our NIRSPEC data, we measured the radial velocity of vB\,10 to 
be $+$34.7\,$\pm$\,1 km\,s$^{-1}$, in very good agreement with the 
various values ($+$35.4, $+$35.2, $+$35.0 km\,s$^{-1}$) found in the 
literature \citep{martin96,mohanty03}. 
We adopted $+$35.0 km\,s$^{-1}$ for the correlation procedure. 
In addition, vB\,10 is known to be one of the slowest rotators 
among ultracool dwarfs 
($v$\,sin\,$i$\,=\,6.5 km\,s$^{-1}$, \citep{mohanty03}). 

We calculated the cross-correlation function in multiple echelle orders 
of the NIRSPEC spectra using the task {\sc fxcor} within {\sc iraf}, 
and measured velocities by fitting a gaussian to the cross-correlation peak. 
We used 4 spectral orders that are free from strong telluric line 
contamination. Two examples of those orders are shown in 
Fig.~\ref{combo}. 
These orders include features due to K\,{\sc i}, 
TiO, FeH, and water vapor. We note that given the good signal-to-noise ratio 
of the spectra and the significant similarity in spectral type 
between LP\,944$-$20 and vB\,10, the peak of the cross-correlation 
function is unambiguously identified. 
The NIRSPEC radial velocities of LP\,944$-$20 shown in Table~\ref{vrad} 
are the mean of the results in the different orders. 
The associated error bars correspond to the standard deviation of the mean.

\section{Discussion}

\begin{figure}
\epsscale{1.0}
\plotone{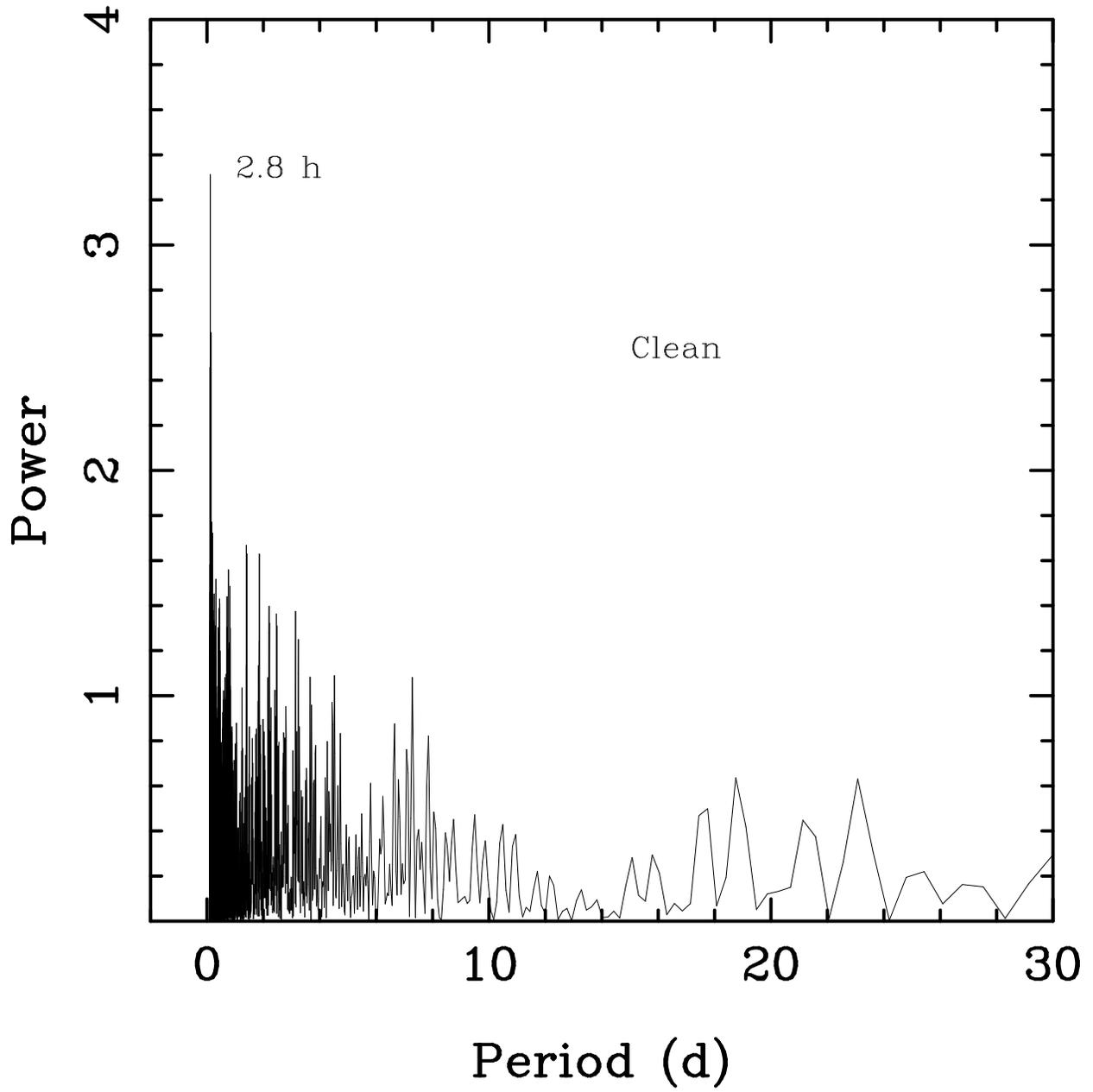}
\caption{\label{periodogram} 
CLEAN periodogram analysis of our UVES RV dataset of LP\,944$-$20. 
5 iterations were used and the gain was 0.1. The highest peak is marked.} 
\end{figure}

\begin{figure}
\plotone{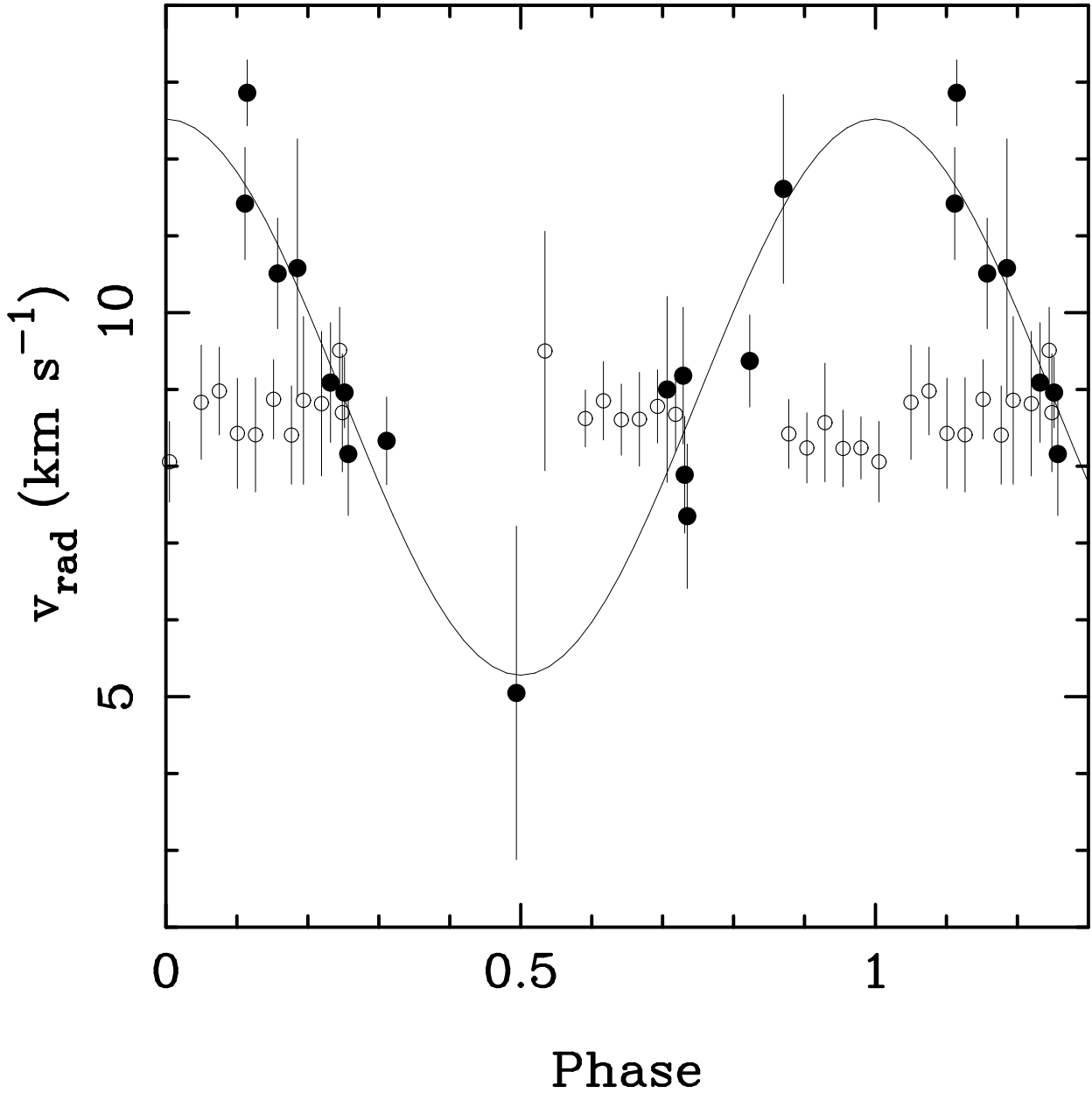}
\caption{\label{phased} 
RV curve of LP\,944$-$20 folded with a period of 3.7 hours  
found in the VLT data. 
A sinusoidal curve provides a good fit for the UVES data (filled circles) 
but not for the NIRSPEC data (open circles). The individual NIRSPEC points 
for each spectrum are shown, whereas nightly averages are given in Table 1.} 
\end{figure}

The UVES RV dataset shows a significant scatter and we performed 
a periodogram analysis of it. The periodogram analysis using 
the SCARGLE and CLEAN programs \citep{scargle,roberts} 
found a most likely periodicity in 
the range 2.5 to 3.7 hours, which is close to the 
expected rotational period of LP\,944$-$20 (vsin $i$=30 km\,s$^{-1}$, P=4.4 hours 
sin~$i$ for R=R$_{\rm Jupiter}$). 
The CLEAN periodogram is shown in (Fig.~\ref{periodogram}). 
We put the  UVES RV data in phase with a period of 3.7 hours  
(Fig.~\ref{phased}) and found a good fit with a sinusoid curve 
with an amplitude of 3.5 km\,s$^{-1}$. 

The NIRSPEC RV data, however, 
do not follow the periodic trend of the UVES data 
(Fig.~\ref{phased}), 
and hence the hypothesis of a planet to explain the UVES RV variations 
can be discarded because the RV signature of a planet should 
not depend on wavelength. In fact, our NIRSPEC data has an rms dispersion 
of 0.36~km\,s$^{-1}$, which is consistent with observational uncertainties. 
Using a mass of 60 Jupiters for LP\,944$-$20, 
we can rule out the presence of any planet with mass larger than 
m~sini$\sim$1.0 Jovian masses and period less than 30 days 
using the NIRSPEC RV dataset. 

 We confirm 
that the RV of  LP\,944$-$20 is variable at optical wavelengths 
as previously reported from UVES observations \citep{guenther03}. 
These authors also noted that the RV variability could not be due to starspots 
because the filling factor would have to be too large. Now that 
we have ruled out the existence of a giant planet, there seems to be only 
one possibility left, which is the presence of inhomogeneous 
surface features that are not like starspots. 
Theoretical models have predicted that there could be dust clouds 
in BDs, and several searches for photometric variability have been 
performed among late-M and L dwarfs 
\citep{bailer01,martin01,gelino02,goldman05}. 
In fact, photometric variability has been detected in  LP\,944$-$20 
at optical wavelengths \citep{tinney99}. On the other hand, 
near-infrared monitoring of about 20 late-M and L dwarfs failed to detect 
any variability \citep{bailer03,koen05}.

LP944-20 and other ultracool dwarfs dramatically violate 
the X-ray/radio-emission relation for normal stars, 
as its quiescent radio emission is  4 to 5 orders of
magnitude brighter than expected from its X-ray emission \citep{berger05}.  
The radio emission observed in some ultracool dwarfs imply a large  
magnetic field strength (over a hundred Gauss 
averaged over the whole surface).  Thus,
while it is now clear that at least some old BDs have strong magnetic
fields, the properties of the resulting coronae are strikingly different
from those of stars.  Possibly, not only the coronae differ from those
of stars but also the structure of the magnetic field itself. As shown
by several authors \citep{chabrier06,dobler06}  
fields of fully convective objects (like BDs)
are expected not to be concentrated in small spots but to be distributed
on a global scale. Additionally, due to the low temperature of old BDs
the degree of ionisation in the atmosphere is very low, leading to a
very low degree of the coupling between the magnetic field and the
atmosphere. The coupling between the gas and the magnetic field is
usually described in terms of the magnetic Reynolds number $R_m=lv/\eta$
(were $l$ is a length scale, $v$ a velocity scale, and $\eta$ the
magnetic diffusivity \citep{priest82}). While for the sun $R_m$ is of the
order of unity, for ultracool dwarfs $R_m$ could be as low as
$10^{-15} -- 10^{-20}$ \citep{meyer99,mohanty02}. 
If this is true, BDs with strong magnetic field
should not have spots, and thus the gas-flow in the atmosphere should
not be effected by the magnetic field. 

We propose here that the 
RV-variability detected with UVES may be due to weather effects, i.e. 
variability of cloud coverage and/or flows, streams or long-lived 
nonmagnetic spots in the upper photosphere 
of this object. The near-infrared observations probe deeper in the 
photosphere, and hence they are less affected by those motions 
or structures. The small amplitude of the photometric and spectroscopic 
variability reported by Tinney \& Trolley (1999) and Guenther \& Wuchterl 
(2003), respectively, indicate that the surface features do not have 
a large contrast with the photosphere, and hence they are not like 
starspots. 

 
Independently from a deep physical understanding of  the cause of 
the RV variations in the optical, we can 
safely conclude on pure empirical grounds  
that future RV-based planet searches in late-M stars and 
BDs will be better 
conducted at near-infrared wavelengths than in the optical because 
the targets are brighter and because the measurements are less 
affected by photospheric noise. This finding encourages the 
use of a new generation of high-resolution near-infrared spectrographs 
\citep{crires,martin05} 
for surveying the planetary population of very low-mass stars 
and BDs.

\acknowledgments

Financial support was provided by the Michelson Science Center,  
NSF research grant AST 02-05862, and Spanish MEC grants AYA2003-05355 
and AYA2005-06453. This paper is 
partly based on observations obtained at the European Southern
Observatory at Cerro Paranal, Chile in program
67.C-0160(A), 68.C-0063(A), and 072.C-0110(B).
We thank the Keck and VLT Observatories staff for their 
support. The W.M. Keck Observatory is operated as a scientific 
partnership between the California Institute of Technology, 
the University of California, and NASA. The Observatory was made 
possible by the generous financial support of the W.M. Keck Foundation. 
The authors extend special thanks to those of Hawaiian ancestry on 
whose sacred mountain we are privileged to be guests.

\clearpage

\clearpage

\clearpage

\clearpage

\clearpage

\begin{deluxetable}{lccc}
\scriptsize
\tablecaption{\label{vrad} RV data for LP\,944$-$20}
\tablewidth{0pt}
\tablehead{\colhead{HJD-2400000.0} & \colhead{RV (km/s)} & 
\colhead{$\sigma$ (km/s)} & Instrument  }
\startdata
51890.80439 & 9.04 & 1.00 & NIRSPEC  \nl
52189.74347 & 9.09 & 0.78 & UVES  \nl
52213.62965 & 9.18 & 0.89 & UVES  \nl
52214.60754 & 11.42 & 0.73 & UVES  \nl
52224.77031 &  8.96 & 0.46 & UVES  \nl
52236.73029 & 12.86 & 0.43 & UVES  \nl
52237.83445 &  8.33 & 0.57 & UVES  \nl
52915.87534 & 10.58 & 1.68 & UVES  \nl
52920.63108 & 10.51 & 0.72 & UVES  \nl
52927.80874 & 11.61 & 1.23 & UVES  \nl
53006.74733 &  7.35 & 0.94 & UVES  \nl
53010.58583 &  7.89 & 0.76 & UVES  \nl 
53010.70228 &  5.05 & 2.17 & UVES  \nl
53017.64839 &  9.00 & 1.21 & UVES  \nl
53022.58255 &  9.37 & 0.60 & UVES  \nl
53030.63830 &  8.16 & 0.80 & UVES  \nl
53669.93870 &  8.25 & 0.43 & NIRSPEC  \nl
53670.97038 &  8.69 & 0.40 & NIRSPEC  \nl
53671.96188 &  8.61 & 0.48 & NIRSPEC  \nl
53672.94703 &  9.50 & 1.56 & NIRSPEC  \nl
53753.70181 &  8.70 & 0.77 & NIRSPEC  \nl
\enddata
\end{deluxetable}

\clearpage

\end{document}